\documentclass[journal,twoside,web]{ieeecolor}
\usepackage{tmi}
\usepackage{cite}
\usepackage{amsmath,amssymb,amsfonts}
\usepackage{algorithmic}
\usepackage{graphicx}
\usepackage{textcomp}
\def\BibTeX{{\rm B\kern-.05em{\sc i\kern-.025em b}\kern-.08em
    T\kern-.1667em\lower.7ex\hbox{E}\kern-.125emX}}
\usepackage{multirow}
\usepackage{subfigure}
\newcommand{\etal}{\textit{et al. }}
\newcommand{\ie}{\emph{i.e. }}

\usepackage{stackengine}
\usepackage{color}
\usepackage{amsmath}
\usepackage[T1]{fontenc}
\usepackage{graphicx}
\usepackage{comment}
\usepackage{amsmath,amssymb} 
\usepackage{color}
\usepackage{epsfig}
\usepackage{amsmath}
\usepackage{amssymb}
\usepackage{subfigure}
\usepackage{amssymb}
\usepackage{mathrsfs}
\usepackage{color}
\usepackage{float}
\usepackage{caption}
\usepackage{colortbl}
\usepackage{arydshln}
\usepackage{multirow}
\usepackage{multicol}
\usepackage{booktabs}

\usepackage{hyperref}
\hypersetup{colorlinks, allcolors=citecolor}
\definecolor{citecolor}{RGB}{34,139,34}

\newcommand\xrowht[2][0]{\addstackgap[.5\dimexpr#2\relax]{\vphantom{#1}}}

\begin{document}
\title{DONet: Dual Objective Networks\\ for Skin Lesion Segmentation}
\author{Yaxiong Wang,
        Yunchao Wei, Xueming Qian,~\IEEEmembership{Member~IEEE, } Li Zhu, and Yi Yang
\thanks{Y. Wang is with the School of Software Engineering, Xi'an Jiaotong University, Xi'an, 710049, China. He is now a visiting Ph.D student at ReLER Lab, University of Technology Sydney. (email: wangyx15@stu.xjtu.edu.cn). }
\thanks{Y. Wei is with the Centre for Artificial Intelligence, University of Technology Sydney, Ultimo, NSW 2007, Australia.
 (email: wychao1987@gmail.com). }
\thanks{X. Qian is with the Key Laboratory for Intelligent Networks and Network Security, Ministry of Education, Xi’an Jiaotong University, Xi’an 710049, China, also with the SMILES Laboratory, Xi’an Jiaotong University, Xi’an 710049,China, and also with Zhibian Technology Co. Ltd., Taizhou 317000, China. (email: qianxm@mail.xjtu.edu.cn).}
\thanks{L. Zhu is with the School of Software, Xi’an Jiaotong University, Xi’an 710049, China. (email: zhuli@mail.xjtu.edu.cn).}
\thanks{Yi Yang is with the Centre for Artificial Intelligence, University of Technology Sydney, Ultimo, NSW 2007, Australia. (email: yee.i.yang@gmail.com).}}

\maketitle

\begin{abstract}
Skin lesion segmentation is a crucial step in the computer-aided diagnosis of dermoscopic images. In the last few years, deep learning based semantic segmentation methods have significantly advanced the skin lesion segmentation results. However, the current performance is still unsatisfactory due to some challenging factors such as large variety of lesion scale and ambiguous difference between lesion region and background. In this paper, we propose a simple yet effective framework, named Dual Objective Networks (DONet), to improve the skin lesion segmentation. Our DONet adopts two symmetric decoders to produce different predictions for approaching different objectives. Concretely, 
 the two objectives are actually defined by different loss functions. In this way, the two decoders are encouraged to produce differentiated probability maps to match different optimization targets, resulting in complementary predictions accordingly. The complementary information learned by these two objectives are further aggregated together to make the final prediction, by which the uncertainty existing in segmentation maps can be significantly alleviated. Besides, to address the challenge of large variety of lesion scales and shapes in dermoscopic images, we additionally propose a recurrent context encoding module (RCEM) to model the complex correlation among skin lesions, where the features with different scale contexts
  are efficiently integrated to form a more robust representation. Extensive experiments on two popular benchmarks well demonstrate the effectiveness of the proposed DONet. In particular, our DONet achieves 0.881 and 0.931 dice score on ISIC 2018 and $\text{PH}^2$, respectively. Code will be made public available.
\end{abstract}


\section{Introduction}
\label{sec:introduction}
Medical image segmentation plays a key role in Computer-Aided Diagnosis (CAD) systems, whose aim is to provide doctors with precise interpretation of medical images. Recently, skin cancer segmentation has attracted much attention due to its low survival rate. For example, the five years survival rate of skin melanoma is even less than 15\%~\cite{biDFL,biDFL1}. Owing to the profound significance of medical image segmentation and the complexity associated with manual segmentation, many researchers have dedicated extensive efforts to skin lesion segmentation in the last few decades~\cite{FTL,biDFL,biDFL1,Illumination,TL}. Benefit from the great progress of deep learning, the segmentation methods based on deep convolutional neural network (DCNN) have achieved encouraging performance~\cite{FocusNet,PhiSeg,BoundLoss,Xnet,RASNet,CLCINet,wangx1,wangx2,wangx3}. Many popular architectures like U-net~\cite{Unet}, attnU-net~\cite{AttnUnet} are proposed and have produced promising results on many medical challenges.

\begin{figure}[t]
\setlength{\abovecaptionskip}{0pt} 
\setlength{\belowcaptionskip}{0pt} 
\begin{center}
    \subfigure[]
    {  \begin{minipage}[t]{0.3\linewidth}
    \includegraphics[height=0.7in,width=1.1in]{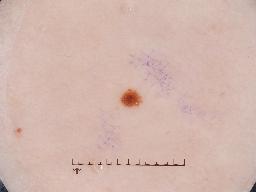}\\
    \vspace{-0.35cm}
    \includegraphics[height=0.7in,width=1.1in]{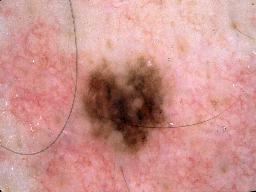}\\
    \vspace{-0.2cm}
    \end{minipage}}
    \subfigure[]{
    \begin{minipage}[t]{0.3\linewidth}
    \includegraphics[height=0.7in,width=1.1in]{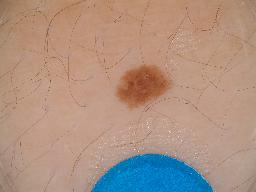}\\
    \vspace{-0.35cm}
    \includegraphics[height=0.7in,width=1.1in]{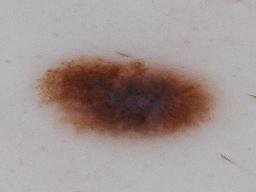}\\
     \vspace{-0.2cm}
    \end{minipage}
    }
    \hspace{-0.25cm}
    \subfigure[]{
    \begin{minipage}[t]{0.3\linewidth}
    \includegraphics[height=0.7in,width=1.1in]{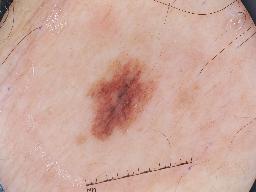}\\
    \vspace{-0.35cm}
    \includegraphics[height=0.7in,width=1.1in]{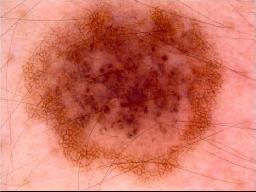}\\
     \vspace{-0.2cm}
    \end{minipage}
    }
\end{center}
\vspace{-0.1cm}
\caption{Six medical images from ISIC 2018 dataset. The large variety of lesion scale and shape poses an important challenge for skin lesion segmentation.}
\label{lesion_scale}
\end{figure}

Besides designing novel losses~\cite{FTL, TL, DiceLoss, BoundLoss} or special architectures~\cite{biDFL, biDFL1, mutiScale, Cascade} to improve medical segmentation, recent researchers find the uncertainty reducing for segmentation map prediction is also extremely useful, especially in clinical applications. Previous works usually integrate multiple maps from different sampling or inference manners to reduce the uncertainty~\cite{Uncertainty,Uncertainty1,Uncertainty2,PhiSeg},  however, these methods all ignore the compensatory feasibility of maps from different optimization objectives. 
Different losses optimize the network from different perspectives, 
we argue that the maps from different objectives could be compensatory. For example, the dice loss encourages the network to produce results with higher dice score coefficient (DSC)~\cite{Illumination}, while the focal tversky loss attempts to achieve the balance between precision and recall~\cite{FTL},  the two maps trained from different losses are biased to respective optimization targets, and they could compensate each other to contribute a more reliable probability map.

However, existing methods simply sum different loss functions and optimize the network to produce only one segmentation map~\cite{ACE,Cascade,FOCUSNET1,biDFL}. The drawbacks of such architectures are two-fold: first, the network with only one expansive path lacks enough degree of freedom to achieve the optimization objectives of different losses simultaneously; second, the only one segmentation map with one-step prediction cannot efficiently interpret the respective targets of multiple losses, since much concession must be made when balancing different targets. To fully exploit the potentiality of different types of losses, we propose to predict separate segmentation maps for different losses and employ two decoders to provide enough degree of freedom to achieve respective optimization objectives, resulting in a simple and effective dual objective network. In particular, a sharing encoder is responsible to extract the feature of the input, and two decoders, which are optimized by different losses, are followed to predict two probability maps, respectively.  The final segmentation map is then jointly decided by integrating the produced two probability maps. Three predicted segmentation maps are supervised by different loss functions, and the network is trained by an end-to-end way. With the designed dual objective architecture, the targets of different losses could be efficiently achieved, what's more important, a joint-decision procedure could be conducted to reduce the uncertainty and produce more reliable segmentation results. 

Beyond the challenge of the uncertainty between the lesion region and its abounding background, the large variety in lesion shape and scale, as shown in Fig.~\ref{lesion_scale}, poses an additional challenge. For example, the area proportion of image lesion in ISIC 2018 dataset~\cite{ISIC2018} ranges from 0.003 to 0.987. Such a large variety requires the lesion segmentation model to be robust to various scale changes. To tackle this challenge, researchers usually  enlarge the receptive field to capture multiple features with different contexts and concatenate them to form a discriminative feature~\cite{ASPP,DenseASPP,biDFL,biDFL1}. However, simple concatenation largely increases the feature dimensionality and fail to investigate the relationship between multiple features with different contexts. To model the contextual relation and distill more beneficial context information from multi-scale context features, we propose a novel recurrent context encoding module (RCEM) to capture a powerful representation, by progressively accumulating the context information with the recurrent neural network. Instead of conducting a simple concatenation, we propose to simulate a feature `zoom-in' procedure using the convolutional long short-term memory (ConvLSTM)~\cite{ConvLSTM} network, which could effectively capture the contextual clues around the lesion. Meanwhile, the encoded knowledge in each time-step is transferred to the expansive path by a multi-scale skip connection, to help the decoders predict accurate probability maps. 

Experiments on two benchmarks demonstrate the effectiveness of our proposed method, we highlight the contributions of this paper as follows:
\begin{itemize}
\item A dual objective network is designed to reduce the uncertainty and achieve a joint decision for the final segmentation map. With the proposed architecture, the probability maps from different optimization aspects work together and compensate each other to jointly predict a more reliable segmentation map.

\item We design a recurrent context encoding module (RCEM), which could model the relation among multi-scale contextual features and efficiently distill more contextual information to address the challenge of variant lesion scales. 
\end{itemize}

\section{Related Work}
\subsection{Uncertainty Estimation}
Uncertainty estimation in medical image segmentation has attracted much interest these years~\cite{Uncertainty2,Uncertainty,Uncertainty1}. Several strategies exist for modeling the ambiguous predictions of neural networks. Kendall \etal use an approximate Bayesian inference mechanism over the network weights to account the uncertainty. Kohl \etal argue that the Bayesian inference may produce samples that vary pixel by pixel, leading to not capture complex correlation structures in the distribution of segmentations~\cite{Uncertainty2}. Therefore, they propose a generative segmentation model based on a combination of U-Net~\cite{Unet} with a conditional variational autoencoder~\cite{CVAE}, which attempts to address the uncertainty by producing an unlimited number of plausible hypotheses~\cite{Uncertainty2}. In ~\cite{UncertainICCV}, the authors reformulate existing single-prediction models as multiple hypothesis prediction (MHP) models for uncertainty reduction. Instead of estimating the uncertainty from multiple hypotheses, Galdran \etal directly introduce an uncertain class and formulate the segmentation task as a multi-class classification problem~\cite{Uncertainty}. A different line of work developed by Baumgartner \etal utilizes the data augmentation technique at inference time, to estimate the uncertainty~\cite{PhiSeg}. Monte Carlo based strategy is also a popular choice. Monte Carlo batch normalization (MCBN) is designed to conduct M forward passes of test sample  at inference time\cite{testAugmentation, MC}. 
Gal \etal design a Monte Carlo Dropout (MCD) that captures predictive uncertainty by turning on the dropout at inference time. Araujo \etal~\cite{DR_uncertain} attempt to measure how much that the decision should be trusted. They design a learning-based diabetic retinopathy grading CAD system to provide a medically interpretable explanation, the designed system could estimate how uncertain that the prediction is. Considering the  intrinsic characteristics of the tracer-kinetic model, Bliesener \etal~\cite{DCE_uncertain} develop a approach for simultaneous estimation of tracer-kinetic parameters and their uncertainty, they train a powerful neural network to estimate the uncertainties for each voxel, which are specific to the patient, exam, and lesion. 
Recently, various deep learning methods take advantage of predictive uncertainty at inference time in several ways, These studies all try to provide a more reliable interpretation of predictions for experts~\cite{Uncertainty3,Uncertainty4,Uncertainty5,Uncertainty6,Uncertainty7}. 

\begin{figure*}
    \centering
    \includegraphics[height=3.2in, width=6.8in]{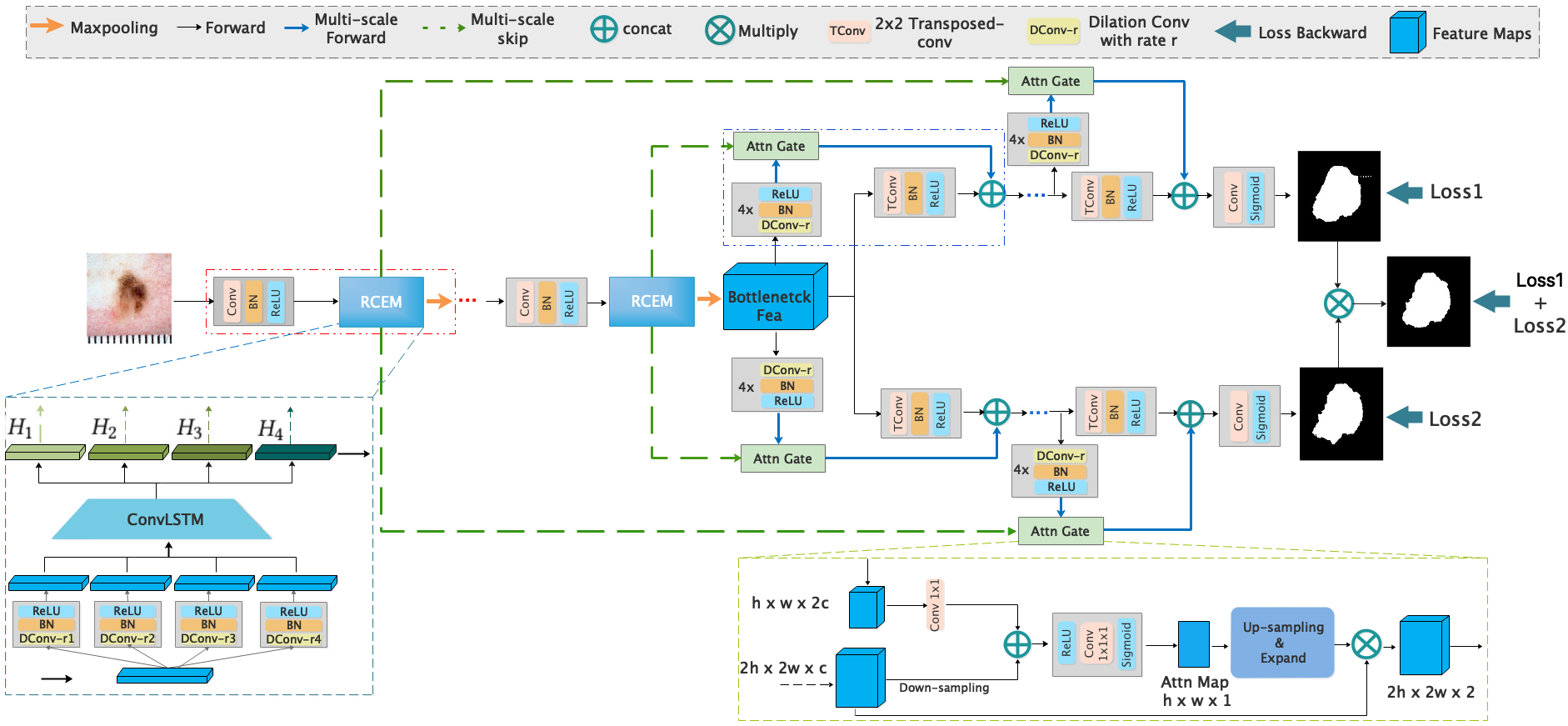}
    \caption{The framework of our proposed dual objective network, where the `...' means repeating the operations in the dashed frame before it. The medical image is first fed through the shared encoder, and two separate decoders are followed to predict two probability maps to meet different optimization objectives. The two probability maps are multiplied to conduct a joint decision for the final segmentation map. Meanwhile, the recurrent context encoding module is plugged in the encoder to capture a powerful contextual feature, and the multi-scale skip is employed to help the decoders produce more reliable results. }
    \label{fig:workflow}
\end{figure*}

\subsection{Contextual Feature Learning}
Fully convolutional neural networks (FCNs) is a popular and fundamental architecture these years, since it could preserve more spatial information and has shown its powerful feature representation ability in many applications, especially in pixel-wise task like segmentation~\cite{FCN1, FCN2, FCN3}. Ronneberger \etal employ an FCNs framework called U-net and design a skip connection mechanism, which could efficiently enhance the feature learning and network training. Inspired by the success of FCNs and U-net, recent medical segmentation approaches commonly design their network based on these two frameworks for contextual feature extraction~\cite{ASPP,biDFL,biDFL1,FocusNet,Unet,DenseASPP,AttnUnet,FCN}. To address the challenge of large variety of lesion region, Chen \etal propose an atrous spatial pyramid pooling (ASPP), which consists of multiple parallel atrous convolutions with different atrous rates and one global average pooling. 
Yang \etal improve the original ASPP and propose a dense ASPP for semantic segmentation~\cite{DenseASPP}. In ~\cite{biDFL}, the authors develop a novel integration way for feature maps with different atrous rates, two types of contextual features are captured from different concatenation directions. Inspired by the ResBlock~\cite{Resnet}, Ibtehaz \etal propose a MultiResBlolck that encodes the contextual features using the residual connection mechanism~\cite{Rethnking}. Different from extracting contextual information from compressed feature maps, recent works attempt to enhance the feature learning using multi-scale inputs~\cite{FTL,Cascade}. Abraham \etal resize the raw image to multiple samples with different resolutions and directly feed them in encoding steps, which aims at recouping the information loss caused by maxpooling operation ~\cite{FTL}. Wang \etal also  utilize the multi-scale images and design a gate-based integration mechanism to efficiently utilize the multi-scale inputs. In~\cite{mutiScale}, Li \etal design a new dense residual deconvolutional network for skin lesion segmentation. The proposed network could capture fine-grained multi-scale features of image by dense deconvolutional layers, chained residual pooling, and an auxiliary supervision mechanism.
Shaban~\cite{CAC} target on the medical image with large resolution, they propose to incorporate large context-aware feature learned from neural network based on the medical images with large resolution, the local feature is frist encoded into high dimensional spacce and then aggregated by considering the spatial organization. Ahn \etal~\cite{unsupervised_context} design an unsupervised approach that uses a multi-layer zero-bias convolutional auto-encoder, meanwhile, a context-based feature augmentation scheme is propose to capture the contextual feature with powerful discriminative power. 

\section{Method}
Fig.~\ref{fig:workflow} exhibits an overview of our proposed method. As shown in Fig.~\ref{fig:workflow}, the designed network is with a horizontal `Y' shape, which comprises of a shared encoder and two separate decoders. The dermoscopic image is first compressed by the encoder, and the bottleneck feature is then fed to the decoders for prediction based on different objectives. The two probability maps from the separate decoders are integrated to conduct a joint optimization for the final prediction. Meanwhile, the designed recurrent context encoding module (RCEM) is plugged in the encoder to learn robust feature representations for lesions of different scales. The details of the dual objective architecture and the RCEM module would be presented in the following.

\begin{figure*}[t]
\setlength{\abovecaptionskip}{0pt} 
\setlength{\belowcaptionskip}{0pt} 
\begin{center}
    \subfigure[Input images]{
    \begin{minipage}[t]{0.16\linewidth}
    \includegraphics[height=0.8in,width=1.25in]{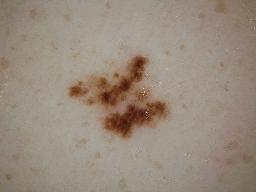}\\
    \vspace{-0.35cm}
     \includegraphics[height=0.8in,width=1.25in]{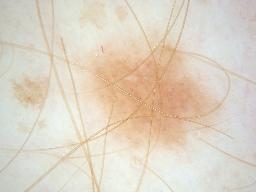}\\
     \vspace{-0.35cm}
    \includegraphics[height=0.8in,width=1.25in]{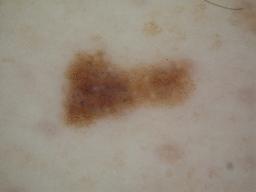}\\
    \vspace{-0.35cm}
    \includegraphics[height=0.8in,width=1.25in]{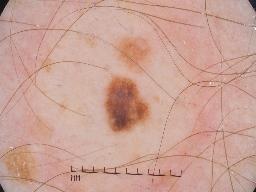}\\
  \vspace{-0.3cm}
    \end{minipage}
    }
    \hspace{0.02cm}
    \subfigure[Prediction1 ($L_1$)]{\begin{minipage}[t]{0.16\linewidth}
    \includegraphics[height=0.8in,width=1.25in]{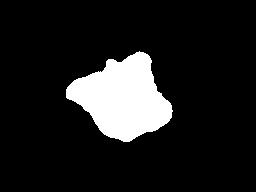}\\
    \vspace{-0.35cm}
    \includegraphics[height=0.8in,width=1.25in]{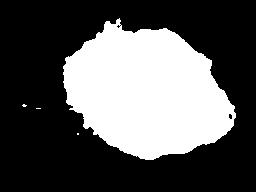}\\
    \vspace{-0.35cm}
    \includegraphics[height=0.8in,width=1.25in]{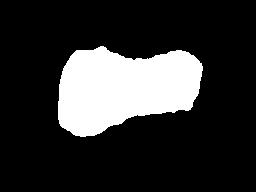}\\
    \vspace{-0.35cm}
    \includegraphics[height=0.8in,width=1.25in]{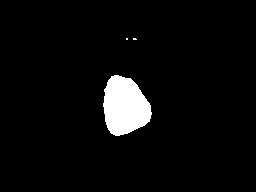}\\
    \vspace{-0.3cm}
    \end{minipage}}
    \hspace{0.02cm}
    \subfigure[Prediction2 ($L_2$)]{
    \begin{minipage}[t]{0.16\linewidth}
    \includegraphics[height=0.8in,width=1.25in]{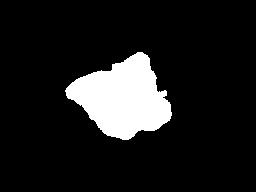}\\
    \vspace{-0.35cm}
   \includegraphics[height=0.8in,width=1.25in]{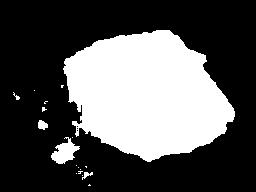}\\
   \vspace{-0.35cm}
   \includegraphics[height=0.8in,width=1.25in]{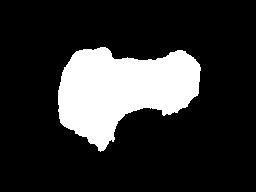}\\
    \vspace{-0.35cm}
    \includegraphics[height=0.8in, width=1.25in]{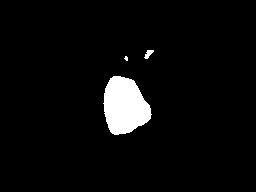}\\
     \vspace{-0.3cm}
    \end{minipage}
    }
    \subfigure[Joint prediction ($L_f$)]{
    \begin{minipage}[t]{0.16\linewidth}
    \includegraphics[height=0.8in,width=1.25in]{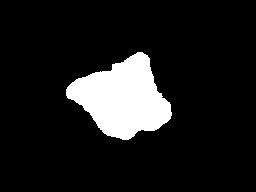}\\
    \vspace{-0.35cm}
    \includegraphics[height=0.8in,width=1.25in]{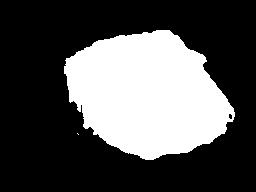}\\
    \vspace{-0.35cm}
    \includegraphics[height=0.8in,width=1.25in]{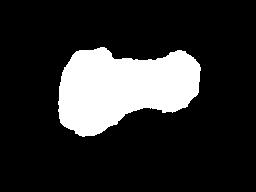}\\
     \vspace{-0.35cm}
     \includegraphics[height=0.8in, width=1.250in]{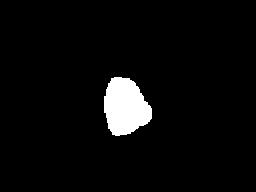}\\
     \vspace{-0.3cm}
    \end{minipage}
    }
    \subfigure[GroundTruth]{
    \begin{minipage}[t]{0.16\linewidth}
    \includegraphics[height=0.8in,width=1.25in]{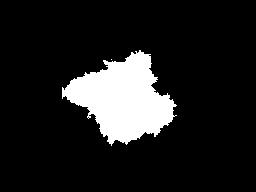}\\
    \vspace{-0.35cm}
      \includegraphics[height=0.8in,width=1.25in]{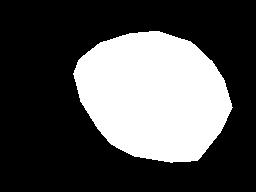}\\
      \vspace{-0.35cm}
     \includegraphics[height=0.8in,width=1.25in]{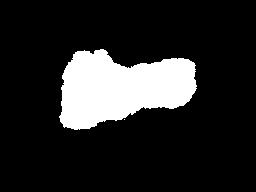}\\
      \vspace{-0.35cm}
      \includegraphics[height=0.8in, width=1.25in]{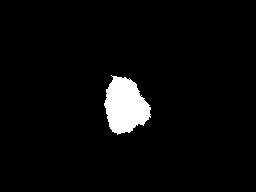}\\
      \vspace{-0.3cm}
    \end{minipage}
    }
\end{center}
\caption{Visual study for our proposed dual objective architecture.}
\label{jpa_dis}
\end{figure*}

\subsection{Dual Objective Network}
An important difference between our network and existing works is that the proposed network is with two decoders, which are with the same architecture but optimized from different objectives. Such a special design is motivated by the intuition that the predictions from different optimization objectives could compensate each other to predict a more reliable segmentation map. With two probability maps meeting different objectives, we can conduct a joint decision to reduce the uncertainty and obtain a more reliable result.

Formally, let $I$ be a dermoscopic image from the training set, $Y$ denotes the corresponding binary mask of $I$, where `1' and `0' refers to the target (lesion) region and the background, respectively. We first feed $I$ through the encoder $E$ to get the bottleneck feature $f$, and then two probability maps are then obtained by the subsequent two separate decoders:
\begin{equation}
\label{d1}
    \hat{Y}^1 = D_1(f; \theta_1),
\end{equation}
\begin{equation}
    \label{d2}
    \hat{Y}^2 = D_2(f; \theta_2),
\end{equation}
where $D$ refers to the decoder, $\theta$ denotes the corresponding parameters, $\hat{Y}^1$ and $\hat{Y}^2$ are the probability maps that have been activated by the sigmoid function.

With the probability maps $\hat{Y}^1$ and $\hat{Y}^2$, the final segmentation map is predicted by conducting an element-wise product:
\begin{equation}
\label{muliply}
\hat{Y} = \hat{Y}^1 \otimes \hat{Y}^2,
\end{equation}
where $\otimes$ is the element-wise product operation. 

The three probability maps are all supervised by the label $Y$ but optimized by three different objectives.
As shown in Eq.~\ref{muliply}, our joint prediction is in fact achieved by a multiplication operation, this procedure is simple but intuitive and efficient. The benefits of this mechanism can be summarized as two-fold:
\begin{itemize}
\item \emph{Effective joint-prediction for the final segmentation map.}

The values of probability maps $\hat{Y}^1$ and $\hat{Y}^2$ both range from 0 to 1, since they have been activated by the sigmoid function. Consequently, 
the multiplication operation in Eq.~\ref{muliply} prompts the two probability maps to discount each other. That is,
the high confidence score in $\hat{Y}$ means that the corresponding predictions in $Y^1$ and $Y^2$ are with more confident predictions, while the low value in $\hat{Y}$ reveals the corresponding two predictions are both low or they strongly disagree with each other. Therefore, for a prediction in $\hat{Y}$, regardless of whether it predicts current pixel is lesion or not, it is with higher probability to be a true prediction comparing to $\hat{Y}^1$ and $\hat{Y}^2$. By simultaneously taking two probability maps into account, we in fact make a joint decision for the final result $\hat{Y}$, which is efficient to reduce the uncertainly for the ultimate probability map. 

\item \emph{Make it possible for the separate decoders to interact.}

The Eq.\ref{muliply} also serves as a hinge to connect the two separate decoders, which could transfer the learned knowledge in the backward procedure and help train a more robust model. Let $L_f$ be the loss function to optimize the gap between the prediction $\hat{Y}$ and the groundtruth label $Y$, then the parameters of two decoders can be updated as follows:
\begin{equation}
\label{update1}
\theta_1^l = \theta_1^{l-1} - \lambda_1 \frac{\partial L_f(\hat{Y}, Y)}{\partial \hat{Y}} \frac{\partial (D_1(I;\theta_1)\otimes D_2(I;\theta_2))}{\partial\theta_1},
\end{equation}
\begin{equation}
\label{update2}
\theta_2^l = \theta_2^{l-1} - \lambda_2 \frac{\partial L_f(\hat{Y}, Y)}{\partial \hat{Y}} \frac{\partial (D_2(I;\theta_2)\otimes D_1(I;\theta1))}{\partial\theta_2},
\end{equation}
where $l$ is the iteration count and $\lambda$ represents the learning rate. Following Eq.~\ref{update1} and Eq.~\ref{update2}, two separate decoders could interact during the training procedure by gradient backward. With such a procedure, the decoders exchange knowledge in each iteration and could benefit each other to produce more reliable predictions.  


\end{itemize}

In our experiments, the two predictions directly from the decoders are trained by different loss functions, $L_1$, $L_2$, respectively. While the joint-prediction $\hat{Y}$ is optimized by the combination of the losses: $L_f=L_1+L_2$. 
A discussion would be presented in our experiment to validate the effectiveness of our dual objective architecture with different types of losses.

Fig.~\ref{jpa_dis} exhibits the visual comparison of the three predictions in the dual objective architecture, where the `Prediction1' and `Prediction2' are the outputs directly from two decoders, the 'Joint prediction' refers to the final prediction from the joint-decision Eq.~\ref{muliply}. From the first row in Fig.~\ref{jpa_dis}, three maps agree with each other on almost all the predictions for a simple test, whose lesion region is clear enough from the background. The second row shows a hard example, it confuses both two decoders, leading to produce many isolated lesions. With our dual objective architecture, the isolated lesion prediction could be suppressed by conducting a comprehensive decision based on the produced two maps. During our experiments, we have two important observations: 1) our joint prediction procedure is good at suppressing the isolated prediction. 2) For simple tests, the three predictions are almost the same, the superiority of our DONet is manifested when encountering difficult samples. Even the two predictions from decoders are both poor, we can still get relatively satisfactory results by the joint prediction procedure, as shown in Fig.~\ref{jpa_dis}.

\subsection{Recurrent Context Encoding Module}
To address the challenge of large variety of lesion scale, we design a Recurrent Context Encoding Module (RCEM), which can gradually capture the contextual features within different scales and integrate them to learn a more robust representation. 
Let $F$ be the feature maps from a certain encoding layer, a group of features with different contexts are then produced by a series of dilated convolution with ascending dilation rates:
\begin{equation}
    \label{dilation_conv}
    F_t = \textit{\text{Conv}}(F, r_t; \delta_t), t\in [1, 2, \cdots, T],
\end{equation}
where the $\textit{\text{Conv}}(F, r;\delta)$ is dilated convolution operation on feature maps $F$ with dilation rate $r$, $\delta$ is the corresponding parameters. As shown in Figure~\ref{fig:workflow}, we follow the setting of~\cite{ASPP} and adopt four convolution kernels (\ie $T=4$) with different dilation rates to capture the contextual features from different receptive fields. Particularly, the feature maps produced by small dilation rates focus on extracting the information of local region from the surrounding pixels, and these maps could preserve the locality of features well. By increasing of the dilation rates, the convolutional operation gets larger receptive fields and could capture more spatial contextual information, making the network harness larger scale context more efficiently.


With the harvested feature maps in $\{F_t\}_{t=1}^T$, a straightforward way for forming scale-friendly feature representation is to simply concatenate them together as adopted by most existing approaches~\cite{biDFL, biDFL1, Cascade, FocusNet}. However, such an operation often largely increases the feature dimensionality and consequently degrades the generalization capability of the feature maps for classification. To tackle this issue, we propose to encode the feature maps equipped with different contextual information to a robust representation by progressively accumulating them with the convolutional LSTM (ConvLSTM)~\cite{ConvLSTM}:
    \begin{align}
    i_t &= \sigma(W_{Fi}* F_t + W_{hi}* H_{t-1}+W_{ci}* C_{t-1}+b_i),\\
    f_t &=\sigma(W_{Ff}* F_t + W_{hf}* H_{t-1}+W_{cf}* C_{t-1}+b_f),\\
    C_t &= f_t \circ C_{t-1} + i_t\tanh(W_{Fc}* F_t + W_{hc}* H_{t-1} + b_c),\\
    O_t &= \sigma(W_{Fo}* F_t + W_{ho}*H_{t-1} + W_{co}\circ C_t + b_c),\\
    H_{t} &= O_t * \tanh(C_t),
    \end{align}
where $W$ and $b$ are the convolutional kernel and bias, respectively, $*$ refers to the convolution and $\circ$ is the Hadamard function~\cite{ConvLSTM}. Such a recurrent encoding can be viewed as a 'zoom-in' procedure, which starts with features with small dilation rate and progressively accumulates the context from the features with larger receptive fields. 
The hidden state $H_t$ captures the context of $\{F_m\}_{m=1}^t$, and the final output $H_T$ efficiently encodes the contextual information of all the feature maps in $\{F_t\}_{t=1}^T$, which is then fed into the following encoding layer for further learning process.

Besides, inspired by the success of skip connection~\cite{Unet}, we further conduct a multi-scale skip to transfer the context knowledge of $\{H_t\}_{t=1}^T$ as the auxiliary information for both decoders. Formally, let $R$ be the output of a decoding layer, whose size is half of $F$ while channel number is double. Symmetrically, a series of features $\{R_t\}_{t=1}^T$ are produced by convolutional operation with dilation rates $\{r_t\}_{t=1}^T$. Instead of directly conducting a concatenation, we employ an attention gate mechanism~\cite{AttnUnet} to suppress the unhelpful information as well as promote the discriminative ones: $A_t = AttnG(H_t, R_t; \eta_t)$, where $\eta_t$ is the respective parameters, the output $A_t$ is with the same shape as $F$  and $AttnG$, whose pipeline is shown in the bottom right of Fig.~\ref{fig:workflow}, is the attention gate mechanism. More details of the $AttnG$ can be found in ~\cite{AttnUnet}. To preserve the original information of the $R$, we directly conduct a transposed convolution with strides 2 on $R$ to produce a feature $Z$, whose shape is the same as $A_t$.  The filtered features $\{A_t\}_{t=1}^T$ are concatenated with $Z$ and fed into the next decoding layer. 

\subsection{Network Training}
Our system produces three kinds of segmentation predictions, \ie, two outputs from the separated decoders and one output from the final joint prediction. These three predictions are optimized by three different types of losses. In our experiments, dice loss~\cite{DiceLoss} and focal tversky loss~\cite{FTL} are applied to optimize the maps directly from the two decoders, respectively.

\noindent \textbf{Dice loss}~\cite{DiceLoss} is proposed to optimize the Dice Score Coefficient (DSC), which is defined as an overlap index between two segmentation maps~\cite{FOCUSNET1}. Specifically, the 2-class DSC loss is defined as:
\begin{equation}
DL = 1 - 2\frac{\sum_{p=1}^N\hat{Y}_pY_p+\epsilon}{\sum_{p=1}^N\hat{Y}_p + Y_p + \epsilon},
\end{equation}
where $\hat{Y}\in [0,1]\text{ and }Y\in \{0, 1\}$ represents the probability map and the ground-truth annotation, respectively, $\epsilon\in \mathcal{R}$ provides numerical stability to prevent division by zero, and $N=W\times H$ is the number of pixels.

\noindent \textbf{Focal tversky loss}~\cite{FTL} improves the tversky loss~\cite{TL} to pursue the balance between precision and recall:
\begin{equation}
\begin{cases}
    &FTL= (1 - TI)^{1/\gamma},\\
    &TI= \frac{\sum_{p=1}^{N}\hat{Y}_pY_{p} + \epsilon}{\sum_{p=1}^{N}\hat{Y}_{p}Y_{p} + \alpha\sum_{p=1}^{N}(1-\hat{Y}_{p})Y_{p}+\beta\sum_{p=1}^{N}\hat{Y}_{p}(1-Y_{p})+\epsilon},
\end{cases}
\end{equation}
where $\gamma,\alpha \text{ and } \beta\in \mathcal{R}$ are the hyperparameters. 

As for the joint prediction, it is trained by the combinations of the two types of losses. We give a discussion on multiple combinations of different losses in our experiment to validate the effectiveness of the proposed dual objective architecture. When the training done, the joint-prediction map $\hat{Y}$ is taken as our final segmentation map in testing stage.
\setlength{\tabcolsep}{2.7mm}{
\begin{table*}
\centering
  \caption{Performance Comparison on ISIC 2018 dataset, the best results are in bold.}
  \vspace{-0.2cm}
  \label{ISICPM}
  \begin{tabular}{lccccc}
  \toprule
     Methods &DSC &JI &Recall  &Precision & Accuracy \\
  \midrule
    FCN~\cite{FCN}& 0.786 &0.701 & 0.897 & 0.718 &0.901 \\
    U-Net~\cite{Unet} & 0.816  &0.727 & 0.904 & 0.740 &0.919 \\
    BCDU-Net~\cite{BCD-Unet} & 0.864 &0.767& 0.927 & 0.758 &0.934 \\
    U-Net++~\cite{unet++} & 0.844 &0.744 &0.889 & 0.752 &0.925 \\
    Attn U-Net~\cite{AttnUnet} &  0.874 & 0.781 &0.929 &0.753 &0.933 \\
    FocusNet~\cite{FocusNet} &0.868 &0.775 &\textbf{0.933} & 0.781 &0.940 \\
    FTL~\cite{FTL} &0.856$\pm$0.007 &0.786$\pm$0.008 &0.897$\pm$0.002 & 0.858$\pm$0.002  &0.945$\pm$0.002\\
    \midrule
    Ours &\textbf{0.881$\pm$0.002} &\textbf{0.806$\pm$0.001} &0.905$\pm$0.002 &\textbf{0.894$\pm$0.002}  & \textbf{0.950$\pm$0.001} \\
  \bottomrule
\end{tabular}
\end{table*}
}
\setlength{\tabcolsep}{3mm}{
\begin{table*}
\centering
  \caption{Performance Comparison on $\text{PH}^2$ dataset, the best results are in bold.}
  \vspace{-0.2cm}
  \label{PH2PM}
  \begin{tabular}{lccccc}
  \toprule
     Methods &DSC &JI &Recall  &Precision & Accuracy \\
  \midrule
    FCN~\cite{FCN}& 0.894 &0.822 & 0.931 & 0.930 &0.935  \\
    U-Net~\cite{Unet} & 0.876 &0.780 & - & - & -\\
    MFCN~\cite{MFCN} & 0.907 &0.840  & 0.949 & 0.940& 0.942\\
    FrCN~\cite{FrFCN} & 0.917  &0.848 & - & - & -\\
    DSL~\cite{DSL} & 0.921 &0.859 & \textbf{0.962} &0.941 &\textbf{0.953} \\
    FTL~\cite{FTL} &0.904$\pm$0.007 &0.825$\pm$0.019  &0.896$\pm$0.021 & 0.924$\pm$0.018 & 0.929$\pm$0.004\\
    \midrule
    Ours &\textbf{0.931$\pm$0.001} & \textbf{0.873$\pm$0.008} &0.936$\pm$0.014 &\textbf{0.945$\pm$0.011} &0.946$\pm$0.007\\
  \bottomrule
\end{tabular}
\end{table*}
}

\section{Experiments}
\subsection{Dataset}
We evaluate the performance of our proposed network on two public benchmarks, \ie, ISIC 2018~\cite{ISIC2018} and $\text{PH}^2$~\cite{PH2}. 

\noindent \textbf{ISIC 2018} was published by the International Skin
Imaging Collaboration (ISIC) as a large-scale dataset of dermoscopy images in 2018. It contains 2594 RGB color images in total and has
become a major benchmark for the evaluation of medical image algorithms. Following Abraham \etal~\cite{FTL}, the dataset is resampled to 192 x 256 pixels with 75-25 train-test split. The training data consists of the raw images and corresponding ground truth annotations. Among training samples, 15\% are randomly selected for validation.

\noindent \textbf{$\text{PH}^2$} is a small dataset, and only contains a total of 200 dermoscopic images of common nevi, atypical nevi, and melanomas, along with their lesion segmentations annotated by an expert dermatologist: 0 for background and 1 for lesion region. They are 8-bit RGB color images with a resolution of 768$\times$560 pixels, which is resized to 256$\times$256 in our experiments. To obtain reliable performance for this dataset, 100 images are used for testing, and 80 images are taken as training data, 20 samples are for validation.

\subsection{Implementation Details}
We employ an U-Net style architecture as the backbone of our DONet. In the encoder part, the input image is downsampled four times by a group of sequential operations: convolution, batch normalization, ReLU activation, and maxpooling. Our proposed RCEM is plugged after each ReLU activation. Two decoders with the same configure are followed to build respective segmentation maps by a series of operations: transposed convolution with strides 2, batch normalization and ReLU activation. In the RCEM, dilation convolution with four dilation rates, \ie, 1, 2, 4, 8, are employed to produce features with different receptive fields. To form a strong baseline, the pyramid inputs~\cite{FTL} are also introduced in the encoding path to recoup more spatial information. 
Following Abraham \etal~\cite{FTL}, the hyperparamters, $\alpha, \beta, \gamma$, for focal tversky loss are fixed as 0.7, 0.3 and 0.75, respectively. 
The final joint prediction map is treated as our final probability map, which is further binarized using threshold 0.5 to get the final segmentation result. 

Our network is trained from an initial learning rate 0.01 for 80 epochs with batch size 8, the learning rate is discounted by 10 for every 40 epochs. To present a fair evaluation of our proposed framework, we do not augment the ISIC 2018 dataset or incorporate any transfer learning. For $\text{PH}^2$ dataset, since this dataset is fairly small, it is hard to train a reliable model only using the original samples. Therefore, we conduct a data augmentation including three random operations: rotation within -20 to 20 degree, horizontal flip and crop, to prevent the network from overfitting. All experiments on both datasets are repeated five times to get credible performance, we report the average performance of all evaluation criteria.

\subsection{Evaluation Criteria}
Five widely used criteria are employed to evaluate the performance including Dice Score Coefficient (DSC)~\cite{Illumination}, Jaccard Index (JI)~\cite{Cascade}, Recall, Precision and Accuracy. The details are as follows:
\begin{align*}
    &\text{DSC} = 2\frac{|GT\cap SR|}{|GT|+|SR|}, \qquad\quad\text{JI} = \frac{|GT\cap SR|}{|GT\cup SR|} \\
    &\text{Recall} = \frac{TP}{TP +FN}, \qquad\quad\text{Precision} = \frac{TP}{TP+FP} \\
    &\text{Accuracy} = \frac{TP + TN}{TP+TN+FP+FN}
\end{align*}
where $GT$ refers to the groundtruth annotation and the $SR$ is the binary segmentation result, and $TP, TN, FP, FN$ are True Positive, True Negative, False Positive, and False Negative, respectively.

\begin{figure}[t]
\setlength{\abovecaptionskip}{0pt} 
\setlength{\belowcaptionskip}{0pt} 
\begin{center}
    \subfigure[Performance on `Benign' images]{
    \begin{minipage}[t]{0.85\linewidth}
    \includegraphics[height=1.5in,width=2.8in]{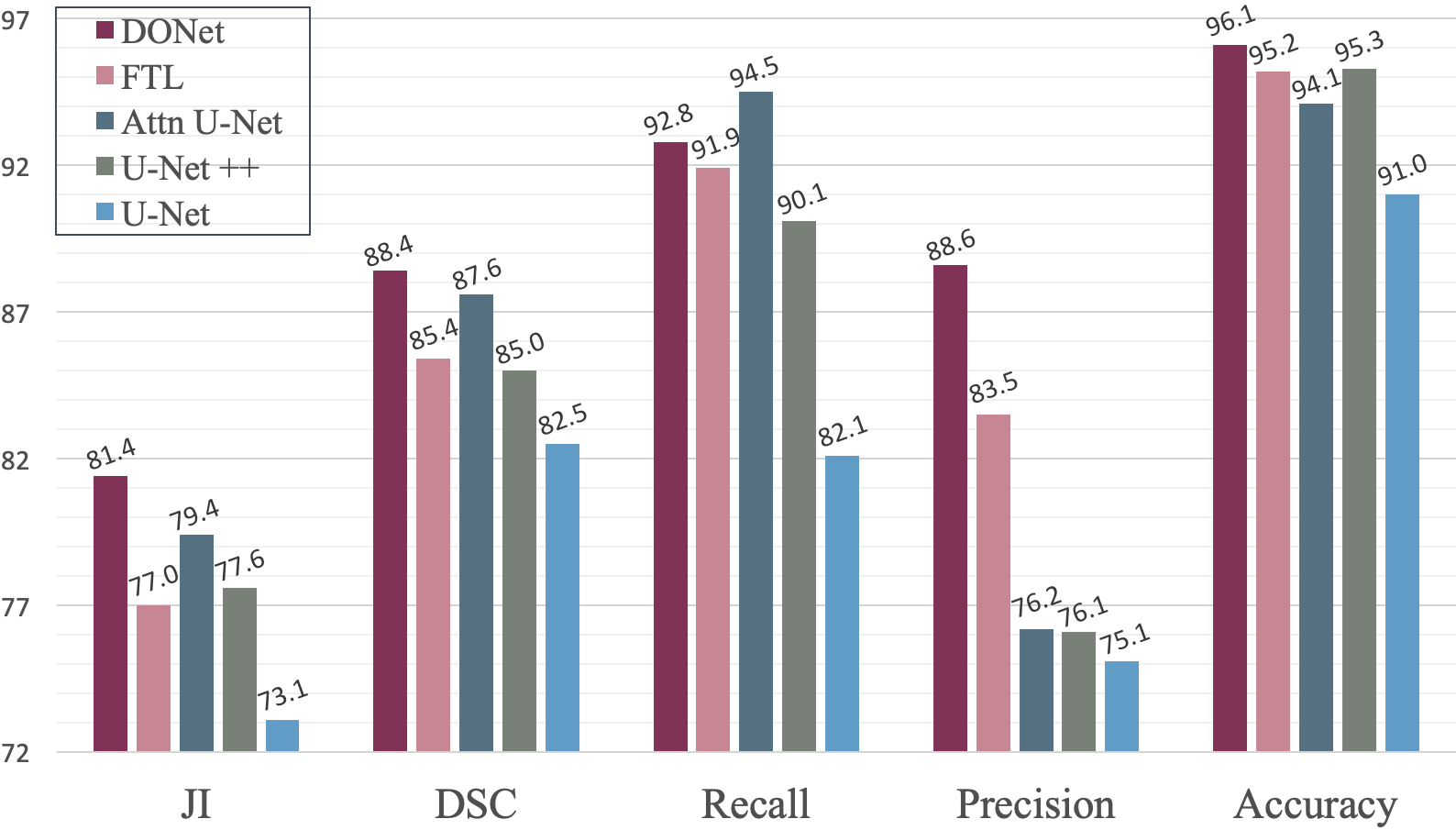}\\
    \vspace{-0.8cm}
    \end{minipage}
    }
    \\
    \subfigure[Performance on 'Malignant' images]{\begin{minipage}[t]{0.85\linewidth}
    \includegraphics[height=1.5in,width=2.8in]{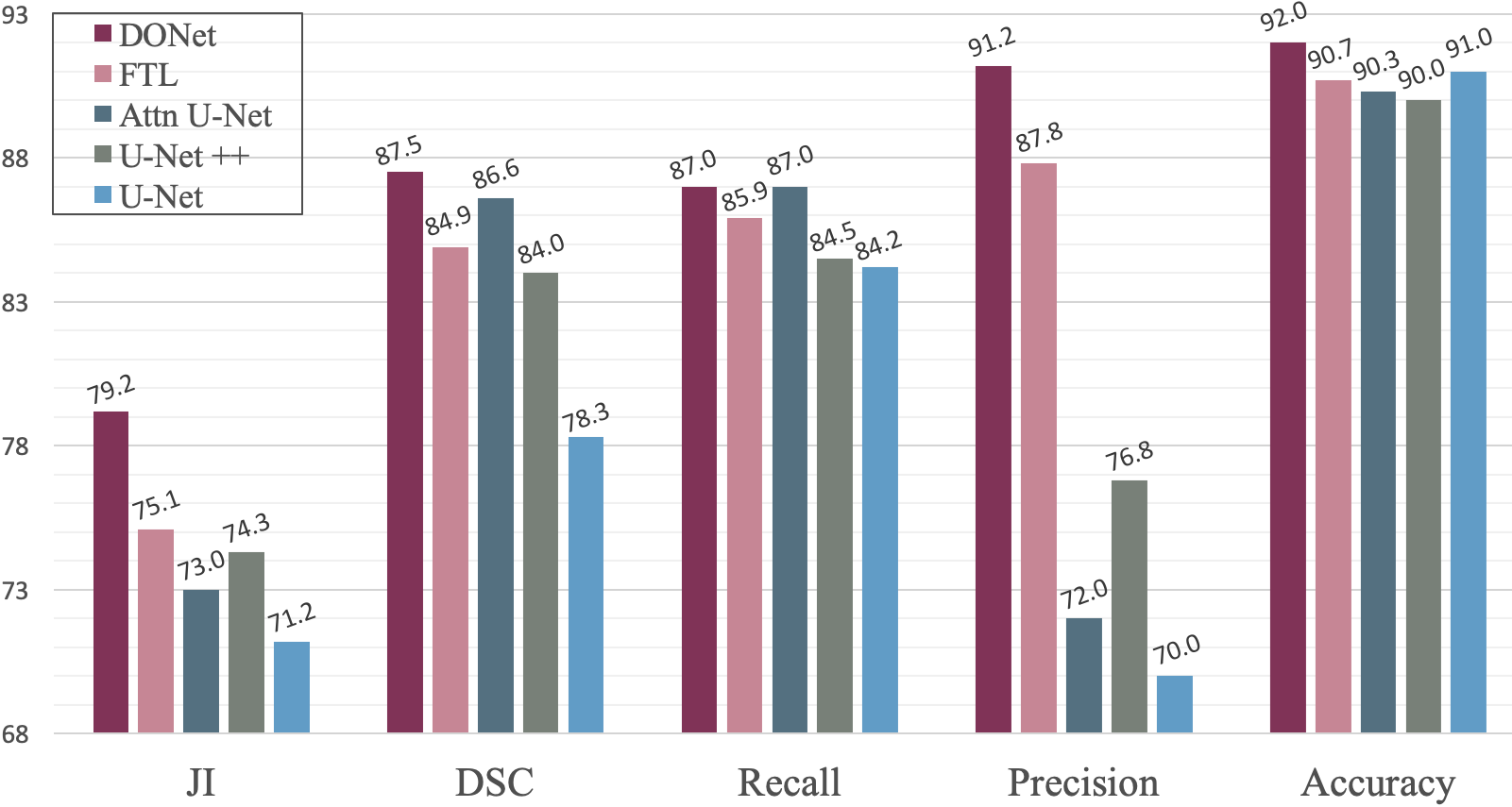}\\
    \vspace{-1cm}
    \end{minipage}}
\end{center}
\caption{Segmentation results on the bengin skin images and the malignant ones for ISIC 2018 dataset. }
\label{class_pm}
\end{figure}

\begin{figure}[t]
\setlength{\abovecaptionskip}{0pt} 
\setlength{\belowcaptionskip}{0pt} 
\centering
\includegraphics[height=2.2in, width=3.2in]{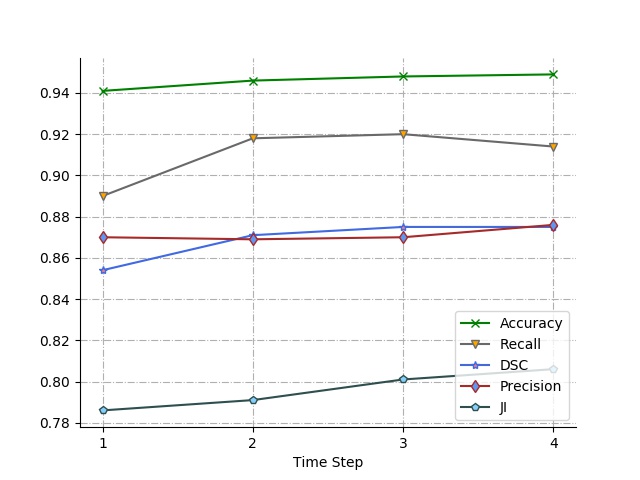}
\caption{The performance tendency when the time step ranges from 1 to 4. }
\label{time_step}
\end{figure}

\subsection{Comparisons with the State-of-the-arts}
Table~\ref{ISICPM} and Table~\ref{PH2PM} show the performance on ISIC 2018 and $\text{PH}^2$ datasets. From these two tables, it is clear that the proposed method outperforms all competing methods. On the ISIC 2018 dataset, our dice score (DSC) could reach 0.881. Under the Precision, the superiority of our DONet is more obvious, since the proposed joint prediction mechanism produces high predictions only if the two predictions from decoders are both confident enough, which encourages the network to produce results with higher precision. The recall of FocusNet~\cite{FocusNet} performs better than our method, however, its poor precision reveals that there are many false positives in its predicted results. While our DONet achieves a good balance between recall and precision. On the $\text{PH}^2$ dataset, our DONet still performs well compared with other approaches. For example, our dice score could reach 0.931, and our Jaccard Index surpasses all competing methods by a large margin. Although the DSL~\cite{DSL} achieves better recall and accuracy, the average performance of our method is more satisfactory.
\begin{figure*}[t]
\setlength{\abovecaptionskip}{0pt} 
\setlength{\belowcaptionskip}{0pt} 
\begin{center}
    \subfigure[Input images]{
    \begin{minipage}[t]{0.13\linewidth}
    \includegraphics[height=0.78in,width=1.06in]{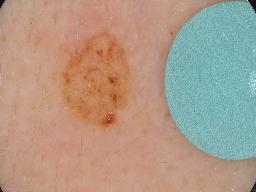}\\
     \vspace{-0.35cm}
   \includegraphics[height=0.78in,width=1.06in]{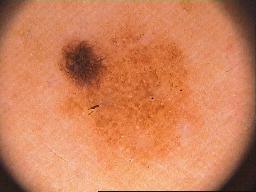}\\
    \vspace{-0.35cm}
   \includegraphics[height=0.78in,width=1.06in]{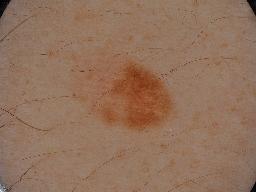}\\
    \vspace{-0.35cm}
    \includegraphics[height=0.78in,width=1.06in]{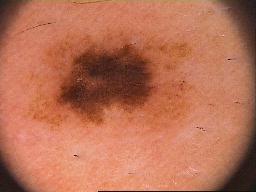}\\
  \vspace{-0.3cm}
    \end{minipage}
    }
    \hspace{0.02cm}
    \subfigure[baseline]{\begin{minipage}[t]{0.13\linewidth}
    \includegraphics[height=0.78in,width=1.06in]{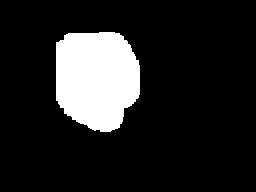}\\
     \vspace{-0.35cm}
   \includegraphics[height=0.78in,width=1.06in]{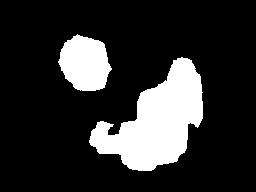}\\
    \vspace{-0.35cm}
   \includegraphics[height=0.78in,width=1.06in]{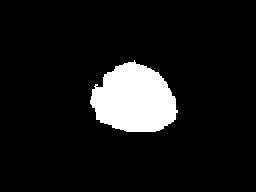}\\
    \vspace{-0.35cm}
  \includegraphics[height=0.78in,width=1.06in]{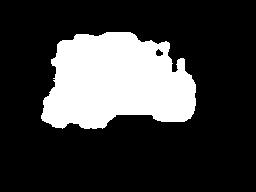}\\

    \vspace{-0.3cm}
    \end{minipage}}
    \hspace{0.02cm}
    \subfigure[baseline+RCEM]{
    \begin{minipage}[t]{0.13\linewidth}
    \includegraphics[height=0.78in,width=1.06in]{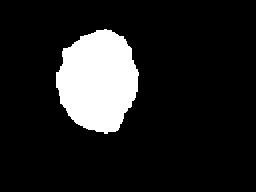}\\
    \vspace{-0.35cm}
 \includegraphics[height=0.78in,width=1.06in]{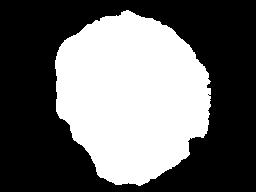}\\
  \vspace{-0.35cm}
 \includegraphics[height=0.78in,width=1.06in]{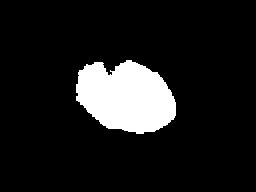}\\
  \vspace{-0.35cm}
 \includegraphics[height=0.78in,width=1.06in]{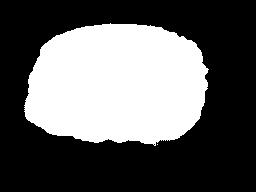}\\
     \vspace{-0.3cm}
    \end{minipage}
    }
    \subfigure[baseline+DOA]{
    \begin{minipage}[t]{0.13\linewidth}
   \includegraphics[height=0.78in,width=1.06in]{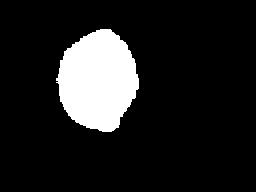}\\
    \vspace{-0.35cm}
   \includegraphics[height=0.78in,width=1.06in]{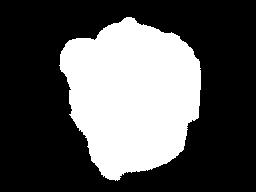}\\
    \vspace{-0.35cm}
   \includegraphics[height=0.78in,width=1.06in]{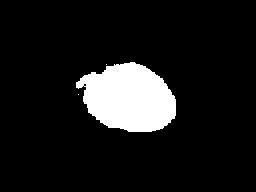}\\
    \vspace{-0.35cm}
   \includegraphics[height=0.78in,width=1.06in]{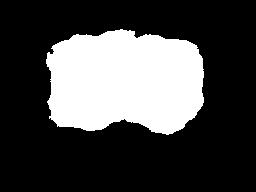}\\
     \vspace{-0.3cm}
    \end{minipage}
    }
    \subfigure[bsl+RCEM+DOA]{
    \begin{minipage}[t]{0.13\linewidth}
  \includegraphics[height=0.78in,width=1.06in]{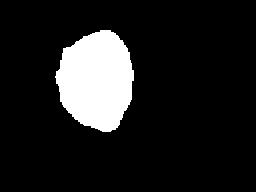}\\
   \vspace{-0.35cm}
  \includegraphics[height=0.78in,width=1.06in]{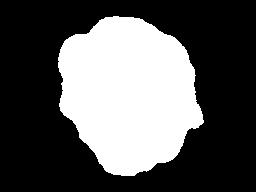}\\
   \vspace{-0.35cm}
  \includegraphics[height=0.78in,width=1.06in]{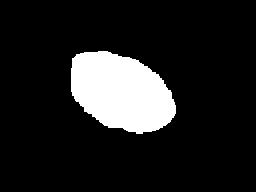}\\
   \vspace{-0.35cm}
  \includegraphics[height=0.78in,width=1.06in]{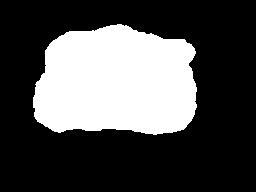}\\
      \vspace{-0.3cm}
    \end{minipage}
    }
       \subfigure[GroundTruth]{
    \begin{minipage}[t]{0.13\linewidth}
   \includegraphics[height=0.78in,width=1.06in]{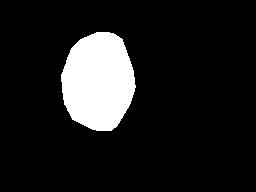}\\
    \vspace{-0.35cm}
  \includegraphics[height=0.78in,width=1.06in]{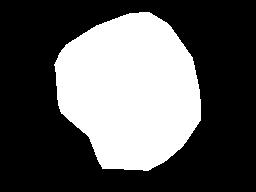}\\
   \vspace{-0.35cm}
  \includegraphics[height=0.78in,width=1.06in]{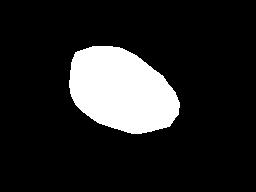}\\
   \vspace{-0.35cm}
   \includegraphics[height=0.78in,width=1.06in]{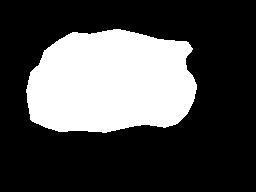}\\
      \vspace{-0.3cm}
    \end{minipage}
    }
\end{center}
\caption{Visual ablation study for the components in our system. where `bsl' means baseline.}
\label{visual_ablation}
\end{figure*}

Fig.~\ref{class_pm} exhibit the performance comparison of five methods on two main categories, \ie the benign lesion image and the malignant ones in ISIC 2018 dataset. From Fig.~\ref{class_pm}, the segmentation results of our DONet also achieves better performance than the competing methods. Although the recall of the Attn-UNet is slightly better than DONet for the benign images, the average performance of our DONet is more outstanding. 

The results from Table~\ref{ISICPM}, Table~\ref{PH2PM} and Fig.~\ref{class_pm} clearly show the superiority of our method, the proposed network could achieve much better average performance under five widely used evaluation criteria, which demonstrate the effectiveness of our method.

\setlength{\tabcolsep}{6.4mm}{
\begin{table*}
\centering
  \caption{Performance comparison with different types of losses, the better results are in bold.}
  \vspace{-0.2cm}
  \label{jha_dis}
  \begin{tabular}{c|c|ccccc}
  \bottomrule\xrowht[()]{9pt}
    Losses & Methods &DSC &Recall  &Precision  \\
   \hline\xrowht[()]{6pt}
    \multirow{2}*{DL + FL} &Baseline &0.857$\pm$0.001  & 0.872$\pm$0.01&0.885$\pm$0.009 \\
                            &DOA &\textbf{0.869$\pm$0.002}  &\textbf{0.874$\pm$0.001} &\textbf{0.905$\pm$0.005}  \\
    \hline\xrowht[()]{6pt}
    \multirow{2}*{DL + TL} &Baseline &0.849 $\pm$0.004 &0.887$\pm$0.01&0.861$\pm$0.014 \\
                            &DOA &\textbf{0.868$\pm$0.002}  &\textbf{0.893$\pm$0.004} &\textbf{0.883$\pm$0.006} \\
    \hline \xrowht[()]{6pt}
    \multirow{2}*{DL + FTL} &Baseline &0.860$\pm$0.003  &0.870$\pm$0.002 &\textbf{0.892$\pm$0.019}  \\
    &DOA  &\textbf{0.871$\pm$0.002} & \textbf{0.893$\pm$0.005} &0.886$\pm$0.008 \\
    \hline\xrowht[()]{6pt}
    \multirow{2}*{FL + TL}&Baseline &0.856$\pm$0.002  &0.885$\pm$0.007 &0.869$\pm$0.009  \\
    &DOA &\textbf{0.870 $\pm$0.002} &\textbf{0.886$\pm$0.003} &\textbf{0.887$\pm$0.005}  \\
    \hline\xrowht[()]{6pt}
    \multirow{2}*{FL + FTL} &Baseline &0.852$\pm$0.004  &\textbf{0.899$\pm$0.011} &0.851$\pm$0.017  \\
    &DOA &\textbf{0.872$\pm$0.001} &0.897$\pm$0.004 &\textbf{0.882$\pm$0.002}  \\
    \hline\xrowht[()]{6pt}
    \multirow{2}*{TL + FTL} &Baseline &0.853$\pm$0.003  &0.891$\pm$0.01 &\textbf{0.859$\pm$0.001} \\
    &DOA &\textbf{0.862$\pm$0.002} &\textbf{0.919$\pm$0.002} &0.846$\pm$0.004  \\
  \bottomrule
\end{tabular}
\end{table*}
}

\subsection{Ablation Study}
To validate the effectiveness and the respective contribution of our proposed dual objective architecture (DOA) and the recurrent context encoding module (RCEM), we thoroughly conduct experiments to show the effectiveness of these two modules. 

\subsubsection{Robustness of DOA with Different Types of Losses}
In our default setting, the dice loss (DL)~\cite{DiceLoss} and the focal tversky loss~\cite{FTL} are used in our dual objective architecture (DOA) to optimize the network. To validate the robustness of the DOA with different types of losses, this subsection,  we thoroughly study the effectiveness of our proposed dual objective architecture by systematically considering four popular losses on ISIC 2018 dataset: Dice Loss (DL)~\cite{DiceLoss}, Focal Loss (FL)~\cite{FocalLoss}, Tversky Loss (TL)~\cite{TL} and Focal Tversky Loss (FTL)~\cite{FTL}. To clearly show the contribution of the DOA, we do not apply the proposed RCEM in this discussion.

Table~\ref{jha_dis} exhibits the performance comparison with different combinations of four losses, where `Baseline' rows indicate the performance of baseline network trained by the sum of corresponding losses in the first column. while the `DOA' rows refer to the results using the proposed dual objective architecture. It is clear that the average performance of DOA is much more outstanding than the baseline methods with different losses. Comparing to our baseline method in Table~\ref{ablation_ISIC}, which is trained by focal tversky loss (FTL) only, the additional introduced dice loss (the sixth row in Table~\ref{jha_dis}) only slightly improves the dice score.  We guess  the reasons for such an observation may stem from two aspects: 1) it is difficult to interpret the different optimization objectives using only one predicted segmentation map, since the 
different losses must make concessions to each other to balance different optimization objectives during training. 2) the commonly used architecture with only one decoder lacks enough degree of freedom to conduct an efficient joint optimization for different types of losses, and even may confuse each other to cause a performance drop, as shown in the twelfth row in Table~\ref{jha_dis} and the second row in Table~\ref{ablation_ISIC}. With our proposed architecture, two decoders are employed to efficiently achieve the objectives of different losses, the joint prediction mechanism integrates the two predicted probability maps to compensate each other and produce a much more reliable segmentation map. The examples shown in Fig.~\ref{jha_dis} intuitively validate the superiority of our dual object architecture.

\setlength{\tabcolsep}{3.6mm}{
\begin{table*}
\centering
  \caption{Contribution of each component in our system on ISIC 2018 dataset, the best results are in bold.}
  \vspace{-0.2cm}
  \label{ablation_ISIC}
  \begin{tabular}{ccc|ccccc}
  \bottomrule\xrowht[()]{9pt}
    & RCEM &DOA &DSC &JI &Recall  &Precision & Accuracy \\
  \hline\xrowht[()]{6pt}
    baseline & & &0.854$\pm$0.004 &0.786$\pm$0.008 &0.890$\pm$0.015 &0.869$\pm$0.02  &0.941$\pm$0.002\\
    baseline &\checkmark & &0.873$\pm$0.003&\textbf{0.822$\pm$0.002} & \textbf{0.913$\pm$0.006} &0.876$\pm$0.015 &0.949$\pm$0.002\\
    baseline & &\checkmark &0.871$\pm$0.002&0.7883$\pm$0.004 & 0.893$\pm$0.005 &0.886$\pm$0.008 &0.948$\pm$0.001\\
    baseline &\checkmark &\checkmark & \textbf{0.881$\pm$0.002} &0.806$\pm$0.012 &0.905$\pm$0.002 &\textbf{0.894$\pm$0.002}  & \textbf{0.950$\pm$0.001} \\
  \bottomrule
\end{tabular}
\end{table*}}

\setlength{\tabcolsep}{3.6mm}{
\begin{table*}
\centering
  \caption{Contribution of each component in our system on $\text{PH}^2$ dataset, the best results are in bold.}
  \vspace{-0.2cm}
  \label{ablation_PH2}
  \begin{tabular}{ccc|ccccc}
  \bottomrule\xrowht[()]{8pt}
    & RCEM &DOA &DSC &JI &Recall  &Precision & Accuracy \\
  \hline\xrowht[()]{6pt}
    baseline & & &0.901$\pm$0.007 &0.825$\pm$0.018 &0.891$\pm$0.019 &0.919$\pm$0.018  &0.930$\pm$0.004\\
    baseline &\checkmark & &0.921$\pm$0.001&0.851$\pm$0.009 & 0.926$\pm$0.005 &0.927$\pm$0.004 &\textbf{0.955$\pm$0.001} \\
    baseline & &\checkmark &0.925$\pm$0.001&0.845$\pm$0.009 & 0.922$\pm$0.008 &0.941$\pm$0.007 & 0.951$\pm$0.002\\
    baseline &\checkmark &\checkmark & \textbf{0.931$\pm$0.0004} & \textbf{0.873$\pm$0.008} &\textbf{0.936$\pm$0.014} &\textbf{0.945$\pm$0.011} &0.946$\pm$0.007 \\
  \bottomrule
\end{tabular}
\end{table*}}

\subsubsection{Discussion of the time step in RCEM}

Different from existing methods that simply concatenate the features with different contextual information, we propose to accumulate the different contextual knowledge by the designed recurrent context encoding module. In our baseline methods, 4-scale (time step) features from the dilation convolution with different dilated rates, \ie 1,2,4,8, are employed to capture the contextual information. In this subsection, we conduct a discussion on ISIC 2018 dataset to show the effect of different time steps. To clearly show the performance change, we also do not employ the proposed dual objective architecture, and directly conduct experiments on the baseline method. 

Fig.~\ref{time_step} shows the performance change when the time-step ranges from 1 to 4, where x-axis indicates the time step, and y-axis shows the value of the criteria. The case that time step=1 means that the RCEM is not employed since there is only a single feature. When a features with larger receptive field is provided, \ie, the case time step=2, the RCEM could be equipped, and a significant performance improvement could be observed. For example, the Dice score is boosted from 0.854 to 0.871 and the Recall could be improved from 0.89 to 0.918, which reveals that accumulating the contextual knowledge using the proposed RCEM is efficient. With the time-step increasing, the performance gain starts to decreasing. Considering the model complexity and the effectiveness, we choose 4 time step in our RCEM module. 

\subsubsection{Respective Contribution of DOA and RCEM to Our System.}

We also conduct experiments on both datasets to clarify the respective contribution of DOA and RCEM, the results on ISIC 2018 and $\text{PH}^2$ datasets are reported in Table~\ref{ablation_ISIC} and Table~\ref{ablation_PH2}, respectively. From Table~\ref{ablation_ISIC}, the dice score of our baseline method is only 0.854. When we plug in the proposed RCEM, the dice score could be improved to 0.873. Our DOA also makes great contribution to our system, and boosts the dice score from 0.854 to 0.871. When we simultaneously equip the RCEM and DOA, the results further get improved, and the final dice score could reach 0.881. From Table~\ref{ablation_PH2}, our RCEM and DOA also make important contribution on $\text{PH}^2$ dataset.  

Fig.~\ref{visual_ablation} shows the visual ablation results. The segmentation maps produced by the baseline method are not satisfactory enough, when the proposed RCEM or DOA are employed, the results get improved. Equipping the two components simultaneously further enhances the performance and could produce more satisfactory results.


\section{conclusion}
This work presents a novel framework for efficient skin lesion segmentation. To produce more credible results, we propose a simple but efficient dual objective architecture, where two separate decoders are employed to provide enough degree of freedom to achieve the different optimization objectives. Consequently, two probability maps meeting different objectives are produced, by conducting a joint prediction based on the two maps, our network could produce much more credible segmentation maps. What's more, a recurrent encoding context module is designed to help capture more powerful contextual features by invoking the convolutional LSTM, and a multi-scale skip mechanism is employed to transfer the learned contextual knowledge to decoding paths for more reliable segmentation map prediction.  Comprehensive experiments on two benchmarks demonstrate that our proposed method is efficient and effective. 


\nocite{*}
\begin{thebibliography}{00}

\bibitem{biDFL} X. Wang, X. Jiang, H. Ding, and J. Liu, ``Bi-Directional Dermoscopic Feature Learning and Multi-Scale Consistent Decision Fusion for Skin Lesion Segmentation", IEEE Trans on Image Processing, vol. 29, pp: 3039-3051, 2020.

\bibitem{biDFL1} L. Ma, and R. C. Staunton, ``Analysis of the contour structural irregularity of skin lesions using wavelet decomposition", Pattern Recognit., vol. 46, no. 1, pp: 98-106, 2013.

\bibitem{FTL} N. Abraham, and N. M. Khan, ``A Novel Focal Tversky Loss Function With Improved Attention U-Net for Lesion Segmentation", ISBI, pp: 683-687, 2019.

\bibitem{FocusNet} C. Kaul, S. Manandhar, and N. E. Pears, ``Focusnet: An Attention-Based Fully Convolutional Network for Medical Image Segmentation", ISBI, pp: 455-458, 2019.

\bibitem{PhiSeg} C. F. Baumgartner \textit{et al.}, ``HiSeg: Capturing Uncertainty in Medical Image Segmentation", MICCAI, pp:119-127, 2019.

\bibitem{BoundLoss} H. Kervadec, J. Bouchtiba, C. Desrosiers, E. Granger, J. Dolz, and I. B. Ayed, ``Boundary loss for highly unbalanced segmentation", MIDL, pp: 285--296, 2019.

\bibitem{Xnet} K. Qi \textit{et al.},
``X-Net: Brain Stroke Lesion Segmentation Based on Depthwise Separable Convolution and Long-Range Dependencies",
MICCAI,
pp: 247--255,
2019.

\bibitem{RASNet} H. Zhang \textit{et al.},
``RSANet: Recurrent Slice-Wise Attention Network for Multiple Sclerosis Lesion Segmentation",
MICCAI,
pp: 411--419,
2019

\bibitem{CLCINet} H. Yang \textit{et al.}, 
``CLCI-Net: Cross-Level Fusion and Context Inference Networks for Lesion
               Segmentation of Chronic Stroke",
MICCAI,
pp: 266--274,
2019.



\bibitem{Rethnking} 
N. Ibtehaz, and M. S. Rahman,
``MultiResUNet : Rethinking the U-Net architecture for multimodal biomedical
               image segmentation",
Neural Networks,
  vol. 121
  pp: 74--87,
2020.

\bibitem{Uncertainty} 
A. Galdran,
               M. I. Meyer,
               P. Costa,
               A. M. Mendon{\c{c}}a, and
               A. Campilho,
``Uncertainty-Aware Artery/Vein Classification on Retinal Images",
ISBI,
  pp: 556--560,
2019.

\bibitem{Unet}
O. Ronneberger,
               P. Fischer, and
               T. Brox,
``U-Net: Convolutional Networks for Biomedical Image Segmentation",
MICCAI,
  pp: 234--241,
2015.

\bibitem{Illumination}
K. Abhishek,
               G. Hamarneh, and
               M. S. Drew,
``Illumination-based Transformations Improve Skin Lesion Segmentation
               in Dermoscopic Images",
{CVPR} Workshops,
pp: 132--3141,
2020.

\bibitem{Uncertainty1}
S. Hu,
               D. E. Worrall,
               S. Knegt,
               B. Veeling,
               H. Huisman, and
               M. Welling,
"Supervised Uncertainty Quantification for Segmentation with Multiple
               Annotations",
MICCAI,
  pp: 137--145,
2019.

\bibitem{Uncertainty2}
S. Kohl \textit{et al.},
``A Probabilistic U-Net for Segmentation of Ambiguous Images",
NIPS,
  pp: 6965--6975,
2018.

\bibitem{DiceLoss}
F. Milletari \textit{et al.},
``V-Net: Fully Convolutional Neural Networks for Volumetric Medical
               Image Segmentation",
3DV,
  pp: 565--571,
2016.

\bibitem{FocalLoss}
T. Lin,
               P. Goyal,
               R. B. Girshick,
               K. He, and
               P. Doll{\'{a}}r,
``Focal Loss for Dense Object Detection",
{IEEE} Trans. Pattern Anal. Mach. Intell.,
  vol. 42,
  no. 2,
  pp: 318--327,
2020.

\bibitem{TL}
S. R. Hashemi \textit{et al.},
``Tversky as a Loss Function for Highly Unbalanced Image Segmentation
               using 3D Fully Convolutional Deep Networks",
CoRR: abs/1803.11078, 2018.

\bibitem{ISIC2018}
N. C. Codella \textit{et al.},
``Skin Lesion Analysis Toward Melanoma Detection 2018: {A} Challenge
               Hosted by the International Skin Imaging Collaboration {(ISIC)}",
CoRR:abs/1902.03368,
2019.

\bibitem{ASPP}
L. Chen,
               G. Papandreou, I. Kokkinos,
               K. Murphy, and
               A. L. Yuille,
``DeepLab: Semantic Image Segmentation with Deep Convolutional Nets,
               Atrous Convolution, and Fully Connected CRFs",
{IEEE} Trans. Pattern Anal. Mach. Intell.,
  vol. 40,
  no. 4,
  pp: 834--848,
2018.

\bibitem{ConvLSTM}
X. Shi \textit{et al.},
``Convolutional {LSTM} Network: {A} Machine Learning Approach for Precipitation
               Nowcasting",
NIPS,
  pp: 802--810,
2015.

\bibitem{CVAE}
D. P. Kingma, and
               M. Welling,
``Auto-Encoding Variational Bayes",
ICLR,
2014.


\bibitem{UncertainICCV}
C. Rupprecht,
               I. Laina,
               R. S. DiPietro, and
               M. Baust,
``Learning in an Uncertain World: Representing Ambiguity Through Multiple
               Hypotheses",
ICCV,
  pp: 3611--3620,
2017.

\bibitem{testAugmentation}
M. S. Ayhan, and P. Berens, 
``Test-time data augmentation for estimation of heteroscedastic aleatoric uncertainty in deep neural networks",
In Medical Imaging with Deep Learning,
2018.

\bibitem{MC}
M. Teye,
               H. Azizpour, and
               K. Smith,
``Bayesian Uncertainty Estimation for Batch Normalized Deep Networks",
ICML,
  pp: 4914--4923,
2018.

\bibitem{MCD}
Y. Gal, and
               Z. Ghahramani,
"Dropout as a Bayesian Approximation: Representing Model Uncertainty
               in Deep Learning",
ICML,
  pp: 1050--1059,
2016.

\bibitem{Uncertainty3}
M. H. Jensen \textit{et al.},
``Improving Uncertainty Estimation in Convolutional Neural Networks
               Using Inter-rater Agreement",
MICCAI,
  pp: 540--548,
2019.

\bibitem{Uncertainty4}
S. Hu,
               D. E. Worrall,
               S. Knegt,
               B. Veeling,
               H. Huisman, and
               M. Welling,
``Supervised Uncertainty Quantification for Segmentation with Multiple
               Annotations",
MICCAI,
  pp: 137--145,
  2019.
  
\bibitem{Uncertainty5}
A. Jungo \textit{et al.},
``On the Effect of Inter-observer Variability for a Reliable Estimation
               of Uncertainty of Medical Image Segmentation",
MICCAI,
pp: 682--690,
2018.


\bibitem{Uncertainty6}
L. Yu,
               S. Wang,
               X. Li,
               C. Fu, and
               P. Heng,
``Uncertainty-Aware Self-ensembling Model for Semi-supervised 3D Left
               Atrium Segmentation",
MICCAI,
  pp: 605--613,
2019.

\bibitem{Uncertainty7}
S. Sedai \textit{et al.},
``Uncertainty Guided Semi-supervised Segmentation of Retinal Layers
               in {OCT} Images",
MICCAI,
pp: 282--290,
2019.

\bibitem{DenseASPP}
M. Yang,
               K. Yu,
               C. Zhang,
               Z. Li, and
               K. Yang,
``DenseASPP for Semantic Segmentation in Street Scenes",
CVPR,
  pp: 3684--3692,
2018.

\bibitem{AttnUnet}
O. Oktay \textit{et al.},
``Attention U-Net: Learning Where to Look for the Pancreas",
CoRR: abs/1804.03999,
2018.

\bibitem{FCN1}
Y. Yuan,
               M. Chao, and
               Y. Lo,
``Automatic Skin Lesion Segmentation Using Deep Fully Convolutional
               Networks With Jaccard Distance",
{IEEE} Trans. Med. Imaging,
  vol.36,
  no. 9,
  pp: 1876--1886,
2017.

\bibitem{FCN2}
L. Bi \textit{et al.},
``Dermoscopic Image Segmentation via Multistage Fully Convolutional
               Networks",
{IEEE} Trans. Biomed. Engineering,
  vol. 64,
  no. 9,
  pp: 2065--2074,
2017.

\bibitem{FCN3}
L. Bi \textit{et al.},
``Step-wise integration of deep class-specific learning for dermoscopic
               image segmentation",
Pattern Recognit.,
  vol. 85,
  pp: 78--89,
 2019.
 
\bibitem{FCN}
E. Shelhamer,
               J. Long, and
               T. Darrell,
``Fully Convolutional Networks for Semantic Segmentation",
{IEEE} Trans. Pattern Anal. Mach. Intell.,
  vol. 39,
  no. 4,
  pp: 640--651,
2017.

\bibitem{Resnet}
K. He,
               X. Zhang,
               S. Ren, and
               J. Sun,
``Deep Residual Learning for Image Recognition",
CVPR,
  pp: 770--778,
2016.

\bibitem{Cascade}
R. Wang,
               S. Chen,
               J. Fan, and
               Y. Li,
``Cascaded Context Enhancement for Automated Skin Lesion Segmentation",
CoRR: abs/2004.08107,
2020.

\bibitem{mutiScale}
H. Li \textit{et al.},
``Dense Deconvolutional Network for Skin Lesion Segmentation",
{IEEE} J. Biomed. Health Informatics,
  vol. 23,
  no. 2,
  pp: 527--537,
2019.

\bibitem{PH2}
T. Mendon{\c{c}}a,
               P. M. Ferreira,
               J. S. Marques, 
               A. R. Mar{\c{c}}al, and
               J. Rozeira,
``PH\({}^{\mbox{2}}\) - {A} dermoscopic image database for research
               and benchmarking",
EMBC,
  pp: 5437--5440,
2013.

\bibitem{RELU}
V. Nair, and
               G. E. Hinton,
``Rectified Linear Units Improve Restricted Boltzmann Machines",
ICML,
  pp: 807--814,
2010.

\bibitem{unet++}
Z. Zhou,
               M. M. Siddiquee,
               N. Tajbakhsh, and
               J. Liang,
``UNet++: {A} Nested U-Net Architecture for Medical Image Segmentation",
MICCAI,
  pp: 3--11,
2018.

\bibitem{BCD-Unet}
R. Azad,
               M. Asadi{-}Aghbolaghi,
               M. Fathy, and
               S. Escalera,
``Bi-Directional ConvLSTM U-Net with Densley Connected Convolutions",
{ICCV} Workshops,
  pp: 406--415,
2019.

\bibitem{MFCN}
L. Bi \textit{et al.},
``Dermoscopic Image Segmentation via Multistage Fully Convolutional
               Networks",
{IEEE} Trans. Biomed. Engineering,
  vol. 64,
  no. 9,
  pp: 2065--2074,
2017.

\bibitem{FrFCN}
M. A. Al{-}masni,
               M. A. Al{-}antari,
               M. Choi,
               S. Han, and
``Skin lesion segmentation in dermoscopy images via deep full resolution
               convolutional networks",
Comput. Methods Programs Biomed.,
  vol. 162,
  pp: 221--231,
2018.

\bibitem{DSL}
L. Bi,
               J. Kim, 
               E. Ahn,
               A. Kumar,
               D. Feng, and
               M. J. Fulham,
``Step-wise integration of deep class-specific learning for dermoscopic
               image segmentation",
Pattern Recognit.,
  vol. 85,
  pp: 78--89,
2019.

\bibitem{ACE}
Y. Zhu \textit{et al.},
``ACE-Net: Biomedical Image Segmentation with Augmented Contracting
               and Expansive Paths",
MICCAI,
  pp: 712--720,
2019.

\bibitem{FOCUSNET1}
Y. Gao \textit{et al.},
``FocusNet: Imbalanced Large and Small Organ Segmentation with an End-to-End
               Deep Neural Network for Head and Neck {CT} Images",
MICCAI,
  pp: 829--838,
2019.


\bibitem{DR_uncertain}
T. Ara{\'{u}}jo \textit{et al.},
``DR{\(\vert\)}GRADUATE: Uncertainty-aware deep learning-based diabetic
               retinopathy grading in eye fundus images",
Medical Image Anal.,
  vol. 63,
  pp: 101715,
 2020.
 
\bibitem{DCE_uncertain}
Y. Bliesener,
               J. Acharya, and
               K. S. Nayak,
``Efficient {DCE-MRI} Parameter and Uncertainty Estimation Using a Neural
               Network",
{IEEE} Trans. Medical Imaging,
  vol. 39,
  no. 5,
  pp: 1712--1723,
2020.

\bibitem{CAC}
M. Shaban \textit{et al.},
``Context-Aware Convolutional Neural Network for Grading of Colorectal
               Cancer Histology Images",
{IEEE} Trans. Medical Imaging,
  vol. 39,
  no. 7,
  pp. 2395--2405,
2020.

\bibitem{unsupervised_context}
E. Ahn,
               A. Kumar,
               M. J. Fulham,
               D. Feng, and
               J. Kim,
``Unsupervised Domain Adaptation to Classify Medical Images Using Zero-Bias
               Convolutional Auto-Encoders and Context-Based Feature Augmentation",
 {IEEE} Trans. Medical Imaging,
  vol. 39,
  no. 7,
  pp: 2385--2394,
2020.

\bibitem{wangx1}
X. Wang \textit{et al.},
``A Weakly-Supervised Framework for {COVID-19} Classification and Lesion
               Localization From Chest {CT}",
{IEEE} Trans. Medical Imaging,
  vol. 39,
  no. 8,
  pp: 2615--2625,
2020.

\bibitem{wangx2}
C. Li,
               X. Wang,
               W. Liu,
               L. J. Latecki, and
               B. Wang,
``Weakly supervised mitosis detection in breast histopathology images
               using concentric loss",
Medical Image Anal.,
  vol. 53,
  pp: 165--178,
2019.

\bibitem{wangx3}
C. Li,
               X. Wang,
               W. Liu, and
               L. J. Latecki,
``DeepMitosis: Mitosis detection via deep detection, verification and
               segmentation networks",
Medical Image Anal.,
  vol. 45,
  pp: 121--133,
2018.

\end{thebibliography}

\end{document}